%% file: main.tex
\documentclass[
twocolumn,
]{ceurart}

\usepackage{listings}
\usepackage{color}
\usepackage{hyperref}
\usepackage{url}
\usepackage{float}
\usepackage[utf8]{inputenc}
\usepackage{amsmath}

\usepackage{amssymb}

\usepackage{enumitem}
\usepackage{cleveref}
\usepackage{natbib}
\usepackage{multirow}
\usepackage{booktabs}       % professional-quality tables
\usepackage{amsfonts}       % blackboard math symbols
\usepackage{nicefrac}       % compact symbols for 1/2, etc.
\usepackage{microtype}      % microtypography
\usepackage{lipsum}
\usepackage{romannum}
\usepackage{wrapfig}

\usepackage{mathtools}
\usepackage{bm}

\usepackage{graphicx}
\usepackage{subcaption}
\usepackage{comment}
\usepackage{algcompatible}
\usepackage{algorithm, algorithmicx}

\newtheorem{definition}{Definition}

\begin{document}

\copyrightyear{2023}
\copyrightclause{Copyright for this paper by its authors.
  Use permitted under Creative Commons License Attribution 4.0
  International (CC BY 4.0).}

\conference{LERI 2023: Workshop on Learning and Evaluating Recommendations with Impressions,
  Sep 19 2023, Singapore}


\title{Impression-Informed Multi-Behavior Recommender System: A Hierarchical Graph Attention Approach}

% \tnotetext[1]{You can use this document as the template for preparing your
%   publication. We recommend using the latest version of the ceurart style.}

%%
% The "author" command and its associated commands are used to define
% the authors and their affiliations.
\author[1]{Dong Li}[%
orcid=0000-0002-2599-6065,
email=dli12@kent.edu,
]
% \cormark[1]
\fnmark[1]
\address[1]{Kent State University}

\author[2]{Divya Bhargavi}[%
orcid=0009-0000-0589-2587,
email=dbharga@amazon.com,
]
\address[2]{Amazon.com, USA}

\author[2]{Vidya Sagar Ravipati}[%
orcid=0009-0009-6909-4985,
email=ravividy@amazon.com,
]

\fntext[1]{This work was completed during the author's internship at Amazon.}

\begin{abstract}
% Implicit feedback has achieved significant success in Recommender Systems. However, multi-behavior interactions between users and items have been overlooked for decades. Traditional implicit recommender systems either homogeneously integrate all types of behaviors (\textit{view}, \textit{add-to-cart}, \textit{buy}, etc) as one (\textit{interacted}) or merely maintain the target behavior (for example, \textit{buy}) by discarding all other auxiliary behaviors. While recent attempts have been made to address this issue, they either optimize only for target behavior, facing data sparsity issues, or ignore hierarchical information between behaviors, leading to sub-optimality. To overcome these limitations, we propose a \textbf{H}ierarchical \textbf{M}ulti-behavior \textbf{G}raph Attention \textbf{N}etwork (HMGN). Our innovative framework applies an attention mechanism to extract information from both inter and intra behaviors, and optimizes a multi-task Hierarchical Bayesian Personalized Ranking (HBPR) criterion. To ensure scalability, we incorporate a multi-behavior sub-graph sampling process. The HMGN can be further enhanced by integrating knowledge metadata and temporal information.  Our models can significantly improve the performance of graph neural network-based methods by up to 64\% in terms of NDCG@100.
While recommender systems have significantly benefited from implicit feedback, they have often missed the nuances of multi-behavior interactions between users and items. Historically, these systems either amalgamated all behaviors, such as \textit{impression} (formerly \textit{view}), \textit{add-to-cart}, and \textit{buy}, under a singular 'interaction' label, or prioritized only the target behavior, often the \textit{buy} action, discarding valuable auxiliary signals. Although recent advancements tried addressing this simplification, they primarily gravitated towards optimizing the target behavior alone, battling with data scarcity. Additionally, they tended to bypass the nuanced hierarchy intrinsic to behaviors. To bridge these gaps, we introduce the \textbf{H}ierarchical \textbf{M}ulti-behavior \textbf{G}raph Attention \textbf{N}etwork (HMGN). This pioneering framework leverages attention mechanisms to discern information from both inter and intra-behaviors while employing a multi-task Hierarchical Bayesian Personalized Ranking (HBPR) for optimization. Recognizing the need for scalability, our approach integrates a specialized multi-behavior sub-graph sampling technique. Moreover, the adaptability of HMGN allows for the seamless inclusion of knowledge metadata and time-series data. Empirical results attest to our model's prowess, registering a notable performance boost of up to 64\% in NDCG@100 metrics over conventional graph neural network methods.
\end{abstract}


\maketitle

\input{text/introduction}
\input{text/problem}
\input{text/framework_new}

\input{text/more}

\input{text/experiment}

\input{text/related}

\input{text/conclusion}

%%
%% Define the bibliography file to be used
%\bibliography{sample-ceur}

\bibliography{ref}

% \appendix


% The sources for the ceur-art style are available via
% \begin{itemize}
% \item \href{https://github.com/yamadharma/ceurart}{GitHub},
% % \item \href{https://www.overleaf.com/project/5e76702c4acae70001d3bc87}{Overleaf},
% \item
%   \href{https://www.overleaf.com/latex/templates/template-for-submissions-to-ceur-workshop-proceedings-ceur-ws-dot-org/pkfscdkgkhcq}{Overleaf
%     template}.
% \end{itemize}

\end{document}

%% file: text/introduction.tex
\section{Introduction}\label{sec:introduction}

Recommender systems are widely employed in online platforms to provide accurate and relevant content to users. As explicit user preferences are often lacking~\cite{als}, implicit feedback has become widely adopted~\cite{als,ncf,rendle@mf,vaecf,Steck_2019,jin@linear,rendle2009bpr}, where the user-item relationship is classified as either \textit{interacted} or \textit{unknown}. However, modern online shopping platforms involve various types of interactions, such as \textit{clicks}, \textit{add-to-cart} actions, and \textit{buy}. Relying solely on binary implicit feedback information (\textit{interacted} or \textit{unknown}) while overlooking this rich multi-behavior information can worsen the cold start problem and exacerbate data sparsity issues~\cite{mbgcn}. Furthermore, considering the naturally existing multi-behavior information in the modern e-commerce ecosystem, it becomes beneficial to differentiate between various user behaviors and optimize for target behaviors, such as \textit{buy}, which align with the ultimate goal of maximizing revenue for businesses. 

Recent years have witnessed the rapid development of multi-behavior recommender systems~\cite{mc-bpr,mb-cross,kgat,xia2021knowledge,chen2021graph,mbgcn,mbgmn,mbht}, due to their ability to supplement and motivate sparse target behavior signals (\textit{buy}) with auxiliary behaviors (\textit{view or what's often termed "impression", favorite, add-to-cart}, etc). Despite these efforts, there exist drawbacks that prohibit these works from fully exploiting the power of multi-behavior information. 

\textbf{Failure to exhaustively utilize and predict auxiliary behaviors.} \cite{mbgcn} constructed a heterogeneous graph convolutional neural network for the multi-behavior recommendation and apply Bayesian Personalized Ranking (BPR), where positive and negative pairs are sampled according to the target behavior. \cite{xia2021knowledge} utilized the graph attention layer together with the item knowledge graph and temporal encoding to empower the learned embeddings to be expressive. All these multi-behavior models \citep{mbgcn,mbgmn,xia2021knowledge} merely optimize for single-behavior (target behavior) despite utilizing multi-behavior information during the model learning stage. Besides, these approaches are unable to predict other auxiliary behaviors thus still suffering from sub-optimal performance and label-sparsity issues. \textbf{Overlooking the hierarchical pattern of multi-behaviors.} \cite{ehcf,chen2021graph} took the backbone of heterogeneous graph convolutional neural network and applied binary Mean Square Error to all behaviors predictions. Despite its efficiency, we argue this method is unable to exploit the inherent relationship between different behaviors as it tries to optimize each behavior to either 1 or 0. \cite{bprh,mc-bpr} gave the behaviors different importance and sample positive and negative items from two different levels of behaviors. These methods did not consider the case where an item would innocently serve as a negative sample while belonging to a higher importance behavior group. More specifically, a user \textit{views} and \textit{buys} an item directly (without \textit{add-to-cart}), would possibly serve as a negative sample when optimizing for \textit{add-to-cart} behavior, which is contrary to the hierarchical pattern between multi-behaviors.

\begin{figure}
\centering
\includegraphics[width=1.\linewidth]{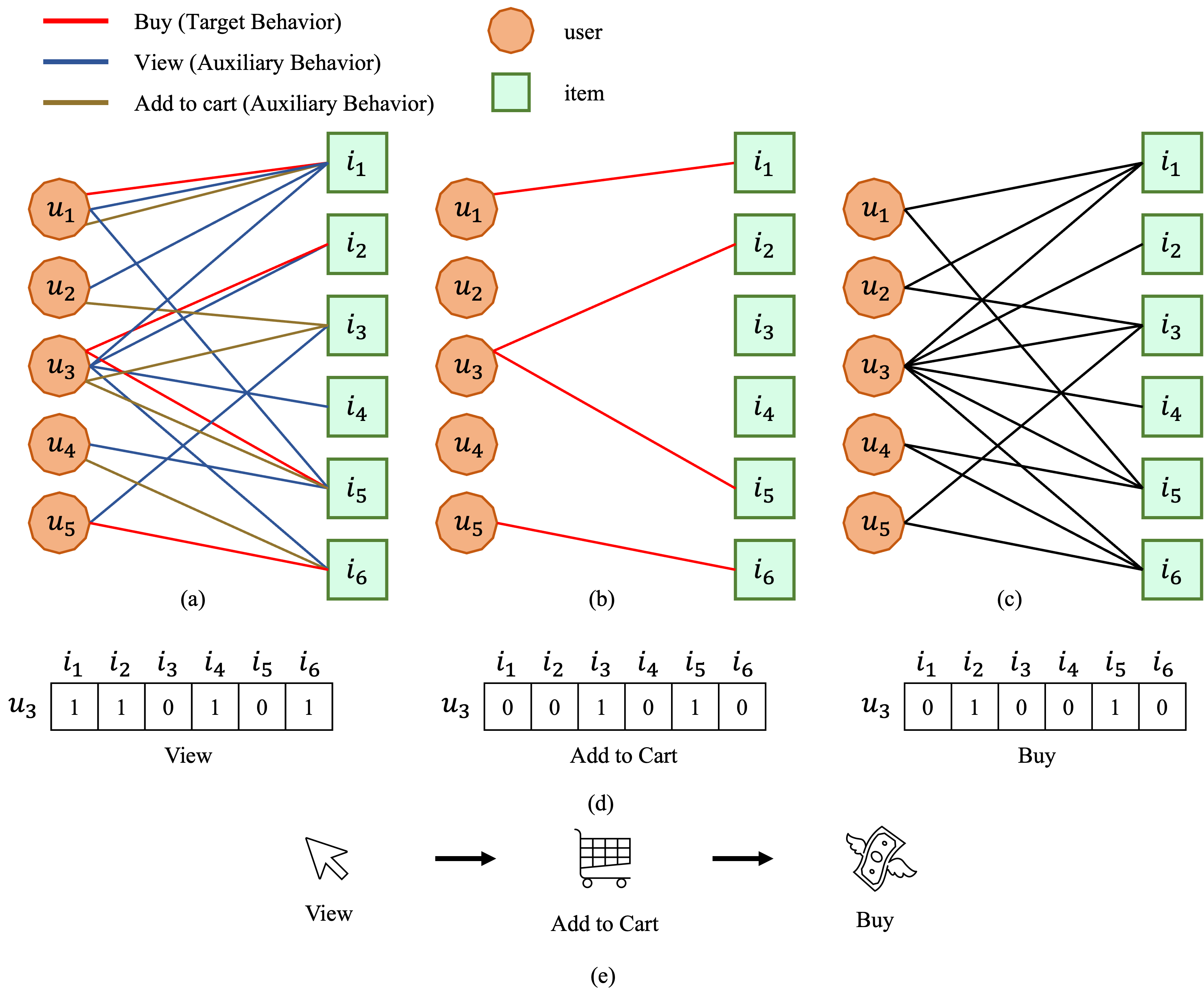}
\caption{An example illustrates the current (multi-behavior) recommender system setting. (a) a multi-behavior example. (b) considering only target behavior. (c) treating all behaviors homogeneously. (d) case of $u_3$. (e) hierarchical shopping pattern. }
\label{fig:single-setting}
% \vspace{-2em}
\end{figure}

To tackle these issues, we aim to leverage the graph attention neural network which enables us to learn and represent individual interaction and optimize a multi-task objective. The challenge is direct utilization of Bayesian Personalized Ranking (BPR) for multi-behavior scenarios can lead to some conflicts. We use an example in \cref{fig:single-setting} (d) to clarify the situation. $u_3$ \textit{add-to-cart} the item $i_3$ and \textit{buy} item $i_2$. To optimize the behavior \textit{add-to-cart}, the BPR criterion would clarify $i_3$ is positive and $i_2$ is negative for $u_3$. We argue that this is not practical since the \textit{buy} behavior over $i_2$ already demonstrates the user's preference. Highly inspired by \cite{mc-bpr,bprh}, we propose a Hierarchical Bayesian Personalized Ranking (HBPR) optimization criterion to deal with the multi-behavior case by taking the hierarchical relations between behaviors into consideration. We propose a multi-behavior ad-hoc graph attention network, aka \textbf{H}ierarchical \textbf{M}ulti-behavior \textbf{G}raph Attention \textbf{N}etwork (HMGN). We list the main contributions of this paper as below:
\begin{enumerate}
\setlength{\itemindent}{-1em}
    \item We explore and benchmark two light yet effective graph attention neural network paradigms targeting on multi-behavior recommendation. Besides target behavior, our model is also able to predict auxiliary behaviors.
    \item We propose HBPR loss criterion dedicated to multi-behavior scenarios. This multi-task optimization framework is crucial for the behavior prediction task.
    \item To scale our model, we extend sub-graph sampling strategy to multi-behavior scenario, avoiding bias in behavior sampling. The corresponding pairwise negative sampling for HBPR loss is refurbished in the sub-graph stage.
    \item Extensive empirical experiments are conducted on two practically processed e-commerce datasets (\textit{Taobao} and \textit{RetailRocket}). The results show that our framework achieves significant improvement over state-of-the-art models (SOTA) (35\% in \textit{RetailRocket} and 64\% in \textit{Taobao}) in terms of offline metric ($NDCG@100$) on target-behavior prediction task.
\end{enumerate}
We organize the rest of this paper in the following way: in \cref{sec:problem}, we formalize the multi-behavior recommendation problem; in \cref{sec:framework}, we detailed clarify the proposed frameworks including model architecture and optimization; in \cref{subsec:sub-sampling}, we introduce the sub-graph sampling as well as incorporation of temporal and knowledge information for multi-behavior recommendation; in \cref{sec:exp}, we conduct extensive empirical experiments; in \cref{sec:related}, we briefly introduce the related works and discuss the resemblance and discrepancy from ours and in \cref{sec:conclusion}, we summarize and conclude the paper.

%% file: text/problem.tex
\section{Experiment Formulation}\label{sec:problem}

Ad Tech and Media customers always encounter label sparsity for single-behavior recommendation systems despite having rich auxiliary information. Incorportating multiple behaviors can also help in the development of a model that can predict a customer's propensity across different stages of purchase funnel.
Additionally, our customers express the need to create a robust decision-making engine that can leverage side information, such as item metadata, to enhance the recommendation process. The work we present here represents a preliminary step towards their long-term goal of building a resilient pipeline for multi-behavior recommendation systems based on Graph Neural Networks (GNNs).

We first introduce the concept of a multi-behavior recommender system in graph terminology.
\noindent \textbf{User-Item Multi-Behavior (Temporal) Bipartite Graph:} the user-item interactions can be treated as a heterogeneous bipartite graph $\mathcal{G} =(\mathcal{E},\mathcal{V})= \{ (u,b_{ui},i)|u\in U, i\in I, b_{ui}\in B\}$ where $U$ is the set of all user nodes, $I$ is the set of all item nodes, $b_{ui}$ indicates user $u$ interacted item $i$ with behavior $b_{ui}$. $B$ is the set of all multi-behavior interaction edges between a user and an item (\textit{view, add-to-cart, favor, buy, etc} for e-commerce datasets). If temporal information is to be considered, the dynamic graph can be represented as:
$\mathcal{G}_t = \{ (u,b_{ui},,t_{ui}, i)|u\in U, i\in I, b_{ui}\in B\}$ where $t_{ui}$ is the timestamp when $u$ interacted with $i$ under $b_{ui}$.
For each user-item node pair $(u,i)$, there can only exist at most one edge per behavior type.

The goal of the multi-behavior recommender system is to utilize all the interaction information (including \textit{view (analogous to "impression" in the context), add-to-cart, favor, buy, etc}) to predict the possible target behavior, typically \textit{buy} behavior.

%% file: text/framework_new.tex
\section{HMGN: Hierarchical Multi-Behavior Graph Attenion Network}\label{sec:framework}
\begin{figure*}
\includegraphics[width=1.\linewidth]{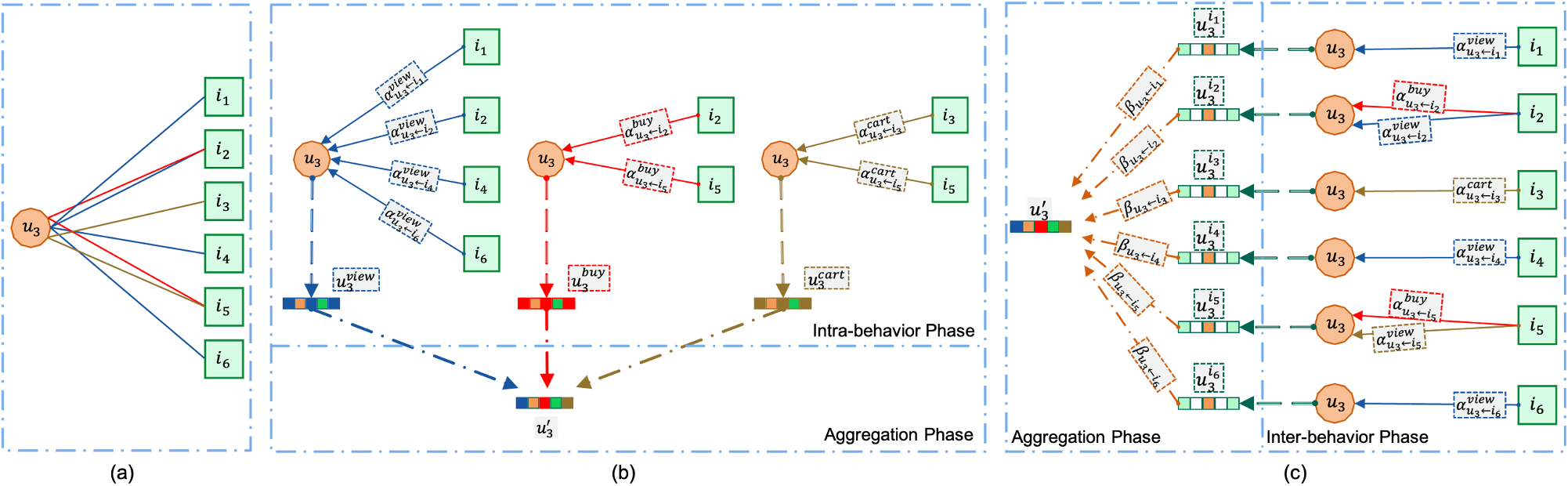}
\caption{Illustration flow of HMGN. (a) shows the example of $u_3$ node with its multi-behavior interacted neighborhood items; (b) is the message propagating flow of HMGN-intra framework;  (c) is the message propagating flow of HMGN-inter framework.}
\label{fig:HMGN}
\end{figure*}

In this section, we elaborate on the details of our proposed framework - \textbf{H}ierarchical \textbf{M}ulti-Behavior \textbf{G}raph Attention \textbf{N}etwork (HMGN) , the model illustration of which is presented in Figure \ref{fig:HMGN}. As we formulated in \cref{sec:problem}, the multi-behavior recommendation data is organized as a heterogeneous bipartite graph. This leads to a different order of information propagation and aggregation process in GNNs based on each behavior which we explore in \textbf{HMGN-intra} and  \textbf{HMGN-inter}. Mathematically, we denote $\mathbf{e}^{(0)}_u$, $\mathbf{e}^{(0)}_i$ as the initialized embedding for user $u$ and item $i$. After propagating through $k$ layers of the graph  attention network, the output of the last layer $\mathbf{e}^{(k)}_u$ and $\mathbf{e}^{(k)}_i$ would be treated as the final representation of user $u$ and item $i$.

\subsection{HMGN-intra}\label{subsec: HMGN-intra}

\subsubsection{Intra-behavior Phase}
Figure \ref{fig:HMGN}(b) illustrates the message propagating and aggregating procedure of the HMGN-intra model for user $u_3$ whose interactions are shown in Figure \ref{fig:HMGN}(a). The heterogeneous user-item bipartite graph is first separated into isolated single-behavior graphs and intra-behavior (message propagating) phase is conducted on these them. We exert the attention mechanism to learn the target node oriented (personalized behavior-specific) representation. Following is the formula of information passing from neighborhood item (source) nodes to its user (target) nodes. This formula is interchangeable between user-to-item and item-to-user information propagation.

\begin{equation}\label{eq:intra-v0}
    \begin{split}
        \alpha^{(l, b)}_{u\leftarrow i}&=\frac{exp \Big( \big(Q^{(b)}\mathbf{e}^{(l)}_u\big)^T\cdot \big(K^{(b)}\mathbf{e}^{(l)}_i\big) \cdot \sqrt{1/d} \Big)}{ \sum\limits_{j\in N^{(b)}_u} exp \Big( \big(Q^{(b)}\mathbf{e}^{(l)}_u\big)^T\cdot \big(K^{(b)}\mathbf{e}^{(l)}_j\big) \cdot \sqrt{1/d} \Big)}\\
        \mathbf{e}^{(l, b)}_u&=\sum\limits_{j\in N^{(b)}_u}{\alpha^{(l, b)}_{u\leftarrow j} \cdot \Big(V^{(b)}\mathbf{e}^{(l)}_j\Big)}
    \end{split}
\end{equation}

where $d$ is hidden dimension, $N^{(b)}_u$ is the set of neighborhood item nodes connected to user $u$ in single-behavior $b$ graph. $Q^{(b)},K^{(b)},V^{(b)}$ are the attention transformation parameters \citep{transformer}  for behavior $b$ and for simplification consideration, we omit the superscripts of layer $l$ from different transformation parameters ($Q,K,V$). Intuitive explanation is: $Q^{(b)}e^{(l)}_u$ would be serve as a \textit{query} transformed from $l$-th layer user representation. \textit{Keys} ($\{K^{(b)}\mathbf{e}^{(l)}_j\}_{j\in N^{(b)}_u}$)  and \textit{values} ($\{V^{(b)}\mathbf{e}^{(l)}_j\}_{j\in N^{(b)}_u}$) are transformed from its neighborhood items. The cross-product attention mechanism (\cref{eq:intra-v0}) is applied to enable the target-oriented representation ($\mathbf{e}^{(l, b)}_u$) learning. Since all the procedure is operated and conducted in each isolated single-behavior graph, the $\mathbf{e}^{(l, b)}_u$ is representative from the perspective of each behavior $b$. The next step is to aggregate each behavior representation.
\subsubsection{Aggregation Phase}
All behavior-specific information (for the target node) would be put together to form a final personalized representation. We are trying to keep the structure as light as possible to maintain the most effective and efficient components for the GNN as LightGCN \citep{lightgcn} demonstrated.
To this end, we take a weighted aggregation approach:
\begin{equation}\label{eq:intra-v1}
    \begin{split}
        \mathbf{e}^{(l+1)}_u = \sum\limits_{b\in B}{w_{u,b}\cdot \mathbf{e}^{(l,b)}_u}
    \end{split}
\end{equation}
In \citep{chen2021graph}, $w_{u,b}$ is set to be manually determined hyper-parameters and is shared between each user and item (as $w_{b}$). A drawback for this setting occurs when a node doesn't exhibit a specific behavior. It leads to an unstable amplitude in $\mathbf{e}^{(l+1)}_u$ with $\mathbf{e}^{(l, b)}_u$ being a zero vector. With our personalized average aggregation, when a user does not exhibit the specific behavior $b$, their weight $w_{u,b}$ shall be set as $0$, and the aggregation operation is performed over the remaining behaviors:
%is: for some node without specific behavior $b$, the zero vector $\mathbf{e}^{(l, b)}_u$ would lead to unstable amplitude in $\mathbf{e}^{(l+1)}_u$. 
\begin{equation}
 w_{u,b} = \begin{cases}
0,&\mathbf{e}^{(l,b)}_u = 0\\
\frac{1}{\sum\limits_{b}^{B}{\delta(\mathbf{e}^{(l,b)}_u \neq 0)}} &\mathbf{e}^{(l,b)}_u \neq 0
\end{cases}
\end{equation}
where $\delta$ is the indicator function, equal to $1$ when input is True and $0$ for False.

\subsection{HMGN-inter}\label{subsec:HMGN-inter}
In \textbf{HMGN-inter} framework, the message would first pass and aggregate across all behaviors within each user-item node pair. We motivate this from real-world patterns where different user/item can exhibit varying levels of multiple behaviors. For example, a particular user can perform more \textit{add-to-cart} actions compared to \textit{buy}. Similarly, a particular item could experience more \textit{page-view} than it has been marked as \textit{favorite}. Thus learning user-item behavior involves understanding these complex dynamics that are user/item centric.

\subsubsection{Inter-behavior Phase}
For each user-item pair, the cross attention attempts to maximize the optimal behavior representation for the target node (user $u_3$ from \cref{fig:HMGN}(a)). 
\begin{equation}\label{eq:inter-v0}
    \begin{split}
        \alpha^{(l, b)}_{u\leftarrow i}&=\frac{exp \Big( \big(Q^{(b)}\mathbf{e}^{(l)}_u\big)^T\cdot \big(K^{(b)}\mathbf{e}^{(l)}_i\big) \cdot \sqrt{1/d} \Big)}{ \sum\limits_{b\in B} exp \Big( \big(Q^{(b)}\mathbf{e}^{(l)}_u\big)^T\cdot \big(K^{(b)}\mathbf{e}^{(l)}_i\big) \cdot \sqrt{1/d} \Big)}\\
       \mathbf{e}^{(l)}_{u\leftarrow i}&=\sum\limits_{b\in B}{\alpha^{(l, b)}_{u\leftarrow i} \cdot \Big(V^{(b)}\mathbf{e}^{(l)}_i\Big)}
    \end{split}
\end{equation}
$Q^{(b)}\mathbf{e}^{(l)}_u$ is the personalized \textit{query} in behavior $b$ space and $\{K^{(b)}\mathbf{e}^{(l)}_j\}_{b}^{B}$, $\{V^{(b)}\mathbf{e}^{(l)}_j\}_{b}^{B}$ are the corresponding \textit{keys} and \textit{values} in each behavior space for item $j$. $\alpha^{(l, b)}_{u\leftarrow i}$ tends to scale the weight that how much information the target node ($u$) obtained from different behavior spaces of the neighborhood $i$ node. And the obtained representation $ \mathbf{e}^{(l)}_{u\leftarrow i}$ contains all the behavior information that passed from node $i$ to node $u$. The next step is aggregating all the information from its neighborhood nodes (belonging to set $N_u$).

\subsubsection{Aggregation Phase}
Unlike GCN \citep{chen2021graph} architecture, we use an attention mechanism to aggregate information from neighboring sources nodes to target nodes \citep{kgat}.
%Different from typical GCN \citep{chen2021graph} in aggregation phase, here we take another attention layer to accumulate the information from neighborhood source nodes to target node \citep{kgat}.
\begin{equation}\label{eq:inter-v1}
    \begin{split}
        \beta^{(l, b)}_{u\leftarrow i}&=\frac{exp \Big( \big(Q\mathbf{e}^{(l)}_u\big)^T\cdot \big(K\mathbf{e}^{(l)}_{u\leftarrow i}\big) \cdot \sqrt{1/d} \Big)}{ \sum\limits_{j\in N_u} exp \Big( \big(Q\mathbf{e}^{(l)}_u\big)^T\cdot \big(K\mathbf{e}^{(l)}_{u\leftarrow j}\big) \cdot \sqrt{1/d} \Big)}\\
       \mathbf{e}^{(l+1)}_u&=\sum\limits_{j\in N_u}{\beta^{(l, b)}_{u\leftarrow j} \cdot \Big(V\mathbf{e}^{(l)}_{u\leftarrow j}\Big)}
    \end{split}
\end{equation}
It is worth noting that the attention parameters in \cref{eq:inter-v1} is different from those in \cref{eq:inter-v0}. $\mathbf{e}^{(l+1)}_u$ would be the output of $l+1$ layer after \textit{inter-behavior phase} and \textit{aggregation} phase.

\subsection{Behavior Preference Modeling}\label{subsec:preference}

After information propagating through $L$ layers of GNN, we obtain final representation $\mathbf{e}^{(L)}_u$ and $\mathbf{e}^{(L)}_i$ for user $u$ and item $i$, respectively. In real-world multi-behavior recommendation, a user would have a \textit{generalization} impression $G$ over an item (\textit{like/dislike}) and also the behavior \textit{specialization} preference $S$, (\textit{view/add-to-cart-buy}, etc). Inspired by this joint relation:
\begin{equation*}
    \begin{split}
        \ln Pr(S,G) = \ln Pr(S|G) + \ln Pr(G)
    \end{split}
\end{equation*}
We propose the following formula that can capture both user item \textit{generalization} preference as well as behavior \textit{specialization} preference:
\begin{equation}\label{eq:beh_alpha}
    \begin{split}
        f(u,b,i)=(\mathbf{e}^{(L)}_u)^T\Big((1-\alpha)B + \alpha I\Big)\mathbf{e}^{(L)}_i
    \end{split}
\end{equation}

where $I$ is identity matrix and $B$ is diagonal matrix representing a behavior type $b$. $\alpha$ is a scalar hyperparameter that balances the trade-off between generalization and specialization. Noting that, if $\alpha = 1$, \cref{eq:beh_alpha} would be the user-item inner product \citep{rendle@mf} - the most common strategy to mimic user-item preference; if $\alpha = 0$, it would collapse to the behavior specialization preference, similar to \citep{chen2021graph}. 

\subsection{HBPR: Hierarchical Bayesian Personalized Ranking}

While recent works \citep{mbgcn,xia2021knowledge,kgat} utilize multi-behavior information in modeling phase, they solely optimize for target behavior (treating only target behavior as positive label) in their objective. We argue that this single-task optimization framework won't be sufficient to exploit the power of auxiliary behavior information. And some multi-task optimization \citep{chen2021graph} ignores the relationship between the behaviors.
Thus, we propose a multi-task optimization criterion - Hierarchical Bayesian Personalized Ranking (HBPR).

The behavior-specific personalized formalization is an extension of \citep{rendle2009bpr} and is defined as $>_{u,b} \subset I^2$, where $i >_{u,b} j $ states as user $u$ prefer item $i$ than item $j$ under the behavior $b$. In multi-behavior scenario,  we maintain $>_{u,b}$ with the  properties of \textit{totality}, \textit{antisymmetry}, \textit{transitivity} (see \citep{rendle2009bpr}) , extended it with extra property - \textit{hierarchy}:
\begin{equation*}
\begin{array}{ll}
% \text{totality}&\forall i, j \in I, b\in B: i \neq j \Rightarrow i>_{u,b} j \vee j>_{u,b} i \\
% \text{antisymmetry}&\forall i, j \in I, b\in B: i>_{u,b} j \wedge j>_{u,b} i \Rightarrow i=j\\
% \text{transitivity}&\forall i, j, k \in I, b\in B: i>_{u,b} j \wedge j>_{u,b} k \Rightarrow i>_{u,b} k \\
&\forall i, j\in I, b_1, b_2\in B: \\
 & i>_{u,b_1} j\wedge j>_{u,b_2} i \wedge b_1 >_p b_2 \Rightarrow i>_{u,b_1} j, i=_{u,b_2} j\\ 
\end{array}
\end{equation*}
where $>_p$ is the heuristically defined priority rank across behaviors.

\noindent\textbf{Intuition of \textit{hierarchy}}: 
Using the same example we talked in \cref{fig:single-setting} in \cref{sec:introduction}:
$u_3$ \textit{add-to-cart} the item $i_3$ and \textit{buy} item $i_2$. We have $i_2>_{u_3, buy} i_3$ and $i_3 >_{u_3, add-to-cart} i_2$. Without defining of \textit{hierarchy}, the BPR criterion would classify $i_3$ is positive and $i_2$ is negative for $u_3$ under behavior \textit{add-to-cart}, $i_3 >_{u_3, add-to-cart} i_2$. This is not quite practical since the typical case is a person \textit{click/view} an item, \textit{favor} it or \textit{add-to-cart}, then \textit{buy} it. Under such a hierarchical assumption, as long as the user \textit{buy} an item, the preference over other behaviors is supposed to be automatically claimed. Following this logic, the e-commerce shopping behaviors priority rank $>_p$ can be set as $b_{\text{buy}} >_p b_{\text{add-to-cart}} >_p b_{\text{view}}$. 

\begin{definition}[Higher priority rank set $b^+$]
Given a behaviors priority rank: $b_0>_p>b_1>_p\cdots >_p b_k$. The higher priority rank set of $b_i (0\le i\le k)$ is denoted as  $b^+_i=\{b_0,b_1,\cdots,b_{i-1}\}$.
\end{definition}
For convenience, we define $I^+_{u,b}\subset I$ as the set of all items that user $u$ interacted with behavior $b$ and $I^-_{u,b} = I\backslash I^+_{u,b}$ as non-interacted or unknown in contrast. Obviously, $I^+_{u,b}\cap I^-_{u,b} = \emptyset$ and $I^+_{u,b}\cup I^-_{u,b} = I$. 

An item from $I^+_{u,b}$ can be treat as a positive sample for user $u$ with the behavior $b$, while the negative one is not directly from $I^-_{u,b}= I\backslash I^+_{u,b}$ according to the requirements of \textit{hierarchy}. 

Equipped with this definition of behavior higher priority rank set, we can determine the negative item sets compatible with \textit{hierarchy} principle:
\begin{equation}\label{eq:compatible_negative}
    \begin{split}
    I^{c,-}_{u,b}=\{i\in I^{-}_{u,b}|i\notin I^{-}_{u,b^\prime}, \forall b^\prime \in b^+\}
    \end{split}
\end{equation}

In BPR \citep{rendle2009bpr}, all user interactions are presumed to be independent of each other. The objective likelihood functions are:

\begin{equation}\label{eq:multi-mle}
    \begin{split}
\mathcal{L}&=\prod_{u,b} p(>_{u,b}|\Theta)\\
&=\prod_{u,b}\prod_{i\in I^+_{u,b}}\prod_{j\in I^{c,-}_{u,b}} p(i>_{u,b} j)\\
    \end{split}
\end{equation}
Substituting the user behavior preference modeling \cref{eq:beh_alpha}, the final objective (with regularization) is:
\begin{equation}\label{eq:final_L}
    \begin{split}
\mathcal{L}=\sum\limits_{u,b}\sum\limits_{i\in I^+_{u,b}}\sum\limits_{j\in I^{c,-}_{u,b}}{
-\log\sigma\Big(f(u,b,i) - f(u,b,j)\Big)
        } + \lambda_\Theta||\Theta||^2
    \end{split}
\end{equation}
% Training triplet $(u,b,i,j)$ in \cref{eq:final_L} is sampled based on \cref{alg:sampling}.

%% file: text/more.tex
\section{Model Scaling and Enhancing}\label{sec:more}

Maintaining a lightweight backbone (\Cref{sec:framework}) has enabled us to explore advanced techniques to enhance and scale our framework to large graphs as discussed in this section below.

\subsection{Multi-Behavior Sub-graph Sampling}\label{subsec:sub-sampling}

\begin{figure*}\label{fig:sub-graph-sampling}
    \centering
    \includegraphics[width=\linewidth]{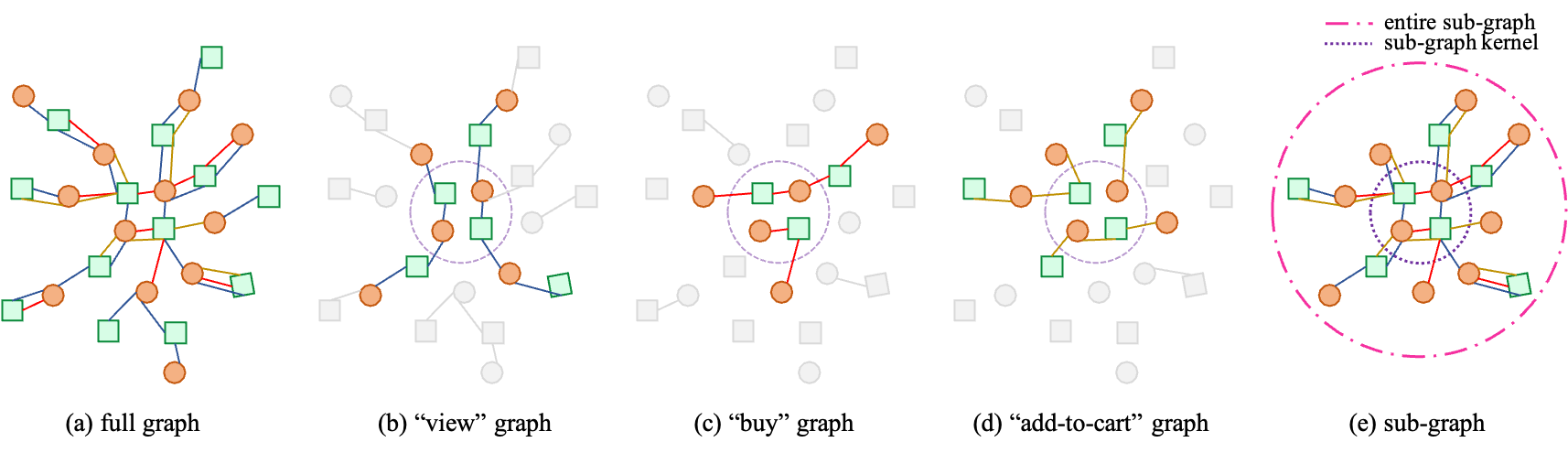}
    \caption{Multi-behavior sub-graph sampling. (a) original full graph; (b)"view"-behavior sub-graph sampling; (c)"buy"-behavior sub-graph sampling; (d)"add-to-cart" sub-graph sampling (e)final sub-graph. For sub-graph BPR optimization, positive user-item pairs are selected from the kernel of sub-graph and negative pairs come from the entire sub-graph.}
    \label{fig:sub-graph-sampling}
\end{figure*}

In real world e-commerce websites, recommender system applications are typically expected to process billions of users and items. Therefore, the scalability of recommendation models is crucial for production deployments. 

While recent works have exhibited impressive success in homogeneous sub-graph sampling strategies (single-behavior) \citep{graphsage,cluster-gcn,graphsaint-ipdps19,layer-sampling}, there is a dearth of literature in the heterogeneous (multi-behavior) sampling realm \citep{hgnn,hgt}. 

In KHGT's \citep{xia2021knowledge} sub-graph wise sampling strategy, the authors first sample target-behavior specific graph to obtain all the nodes. They then form a sub-graph by adding auxiliary behavior edges if present. Further, in the optimization stage, a BPR optimization criterion is applied only to the target behavior. This poses a major drawback - \textit{biased multi-behavior distribution}. Since the nodes in the sub-graph are determined solely by the presence of target behavior edges, the obtained sub-graph has a behavior bias with skewed multi-behavior edge ratios (the target behavior edge counts higher than the other behavior edges).
In this section, we propose a multi-behavior sub-graph sampling strategy (see \Cref{fig:sub-graph-sampling} and \Cref{alg:sub-graph}), which enables tunable behaviors distribution, and the corresponding pairwise HBPR sampling.

At a high level, we sample the sub-graph for each single-behavior graph and obtain the final multi-behavior sub-graph by the union of all the nodes and their connected edges. For each behavior, we first determine a small set of user and item nodes, which we call the \textit{sub-graph kernel} (the inner circle in \Cref{fig:sub-graph-sampling}). Then, we sample up to $k$-hop neighborhood nodes for these kernel nodes. In the HBPR optimization stage, where a pair of nodes (positive and negative item samples for a user) are required, we constrain the positive sample to be derived from our core sub-graph kernel and the negative sample to be derived from rest of the sub-graph. (\Cref{alg:sub-graph})

The justification of why sub-graph kernel is crucial for optimization is as follows. In the original full-size graph, nodes and edges are self-contained (connected), and every node can receive information from all of its neighborhoods  during the graph propagation stage. In the case of sub-graph, since only few nodes are kept, there is an unavoidable loss of knowledge during the information propagation. Especially for those nodes in the "border" of the sub-graph in \Cref{fig:sub-graph-sampling}, they are left with spare neighboring. In contrast, for those nodes in the kernel, almost all the neighbors (dense neighborhoods) would be kept due to our layer-wise sampling strategy. This leads to a more accurate and informative embedding representation for the nodes in the kernel. Thus it would be beneficial to only optimize the kernel edges as positive relations.

\begin{algorithm}[t]
\caption{
Multi-behavior sub-graph HBPR sampling}\label{alg:sub-graph}
\begin{flushleft}
        \textbf{INPUT:} 
        kernel users set ${U}^{k}\subset U$
        \\
        \textbf{OUTPUT:} sub-graph $\mathcal{G}^S=\{(u, b_{ui}, i)|u\in U, i\in I\}$, training set $T^S = \{(u, b, i, j)\}$
\end{flushleft}

\begin{algorithmic}[1]
\FORALL{$b \in \mathcal{B}$}
\STATE construct single-behavior graph $\mathcal{G}_b$
\STATE construct kernel items set $I^k_{b}$ from one-hop neighbors of $U^{k}$
\STATE sampling and obtain single-behavior sub-graph $\mathcal{G}^S_b$
\ENDFOR
\STATE sub-graph $\mathcal{G}^S \gets \{(u,b_{ui},i)| u\in \bigcup_b\mathcal{G}^S_b, i\in \bigcup_b\mathcal{G}^S_b\}$
\STATE sub-graph kernel $\mathcal{G}^k\gets \{(u,b_{ui},i)| u\in U^k, i\in i\in \bigcup_b I^k_b\}$

\STATE Empty dataset $T^S \gets \{\}$
\FORALL{$u \in {U}^k$}
\FORALL{$b \in {B}$} 
    \STATE $I^+_{u,b}\gets \{i | (u,b,i)\in \mathcal{G}^k\}$
    \STATE $I^-_{u,b}\gets \{i\in \mathcal{G}^S |  i\in I\backslash I^+_{u,b}\}$
    \STATE $I^{c,-}_{u,b}\gets\{i\in I^{-}_{u,b}|i\notin I^{-}_{u,b^\prime}, \forall b^\prime \in b^+\}$
    \FORALL{$i\in I^{+}_{u,b}$}
        \STATE $count \gets 0$
        \REPEAT
        \STATE $count \gets count + 1$
        \STATE sampling one negative item $j$ from $I^{c,-}_{u,b}$
         \STATE add $(u,b,i,j)$ to $T^S$
        \UNTIL $count = n_b$
    \ENDFOR
\ENDFOR
\ENDFOR
\end{algorithmic}
\end{algorithm}

\subsection{Temporal Encoding}

Temporal information can be seamlessly incorporated into our framework in a way similar to positional encoding in transformer\citep{transformer,xia2021knowledge}. Formally, we define the temporal representation as a vector $PE_t\in \mathbb{R}^d$ ($d$ is the latent dimension of the model):
\begin{equation}
    \begin{split}
        PE_{t, 2\epsilon}=\sin(\frac{t}{10000^{\frac{2\epsilon}{d}}})\quad  PE_{t, 2\epsilon + 1}=\cos(\frac{t}{10000^{\frac{2\epsilon+1}{d}}})\\
    \end{split}
\end{equation}
where $t$ is the numerated timestamp, $\epsilon\in\{0,1,\dots,d/2-1\}$. 

The temporal vector can be (element-wise) add onto source nodes before the message propagating phase:
\begin{equation}\label{eq:temporal_add}
    \begin{split}
        \mathbf{e}^{(l,t)}_i = \mathbf{e}^{(l)}_i + PE_t
    \end{split}
\end{equation}
The derived temporal representation $\mathbf{e}^{(l,t)}_i$ is used to substitute $\mathbf{e}^{(l)}_i$ in \Cref{eq:intra-v0} and \Cref{eq:inter-v0}. Note here that the temporal information is not feasible for GCN-based models \citep{lightgcn,chen2021graph}, where the message propagation and aggregation is in-batch operated by adjacency (Laplacian) matrix.

\subsection{Knowledge Graph (KG) Enhancing}
We are also able to enhance and boost our framework with metadata. Similar to \citep{kgat,transr}, we leverage the item-meta data and train a separate KG loss. 

\noindent\textbf{Collaborative Knowledge Graph (CKG)} can be seamlessly extended from single-behavior recommendation to multi-behavior recommendation. User-item interactions can be largely divided into multiple triples, $(u, b, i)$ for example, $(u, buy, i)$, $(u, view, j)$, etc. As to the item-meta data, we can also define $(i, r, e)$ where $r\in \mathcal{R}$ is the relation and $e\in \mathcal{E}$ is an entity (item feature), for example, $(i, belongTo, category\ e)$. To this end, the CKG is defined as $\{(h,r,t)|h,t\in U\bigcup I \bigcup \mathcal{E}, r\in B\bigcup \mathcal{R}\}$ The translation principle which optimizes KG by projecting relations and entities into a common semantic space, \citep{transr} is given by:$W_re_h + \mathbf{e}_r\approx W_re_t$, where $\mathbf{e}^r_h$ and  $W_re_t$ are the representation of the head and tail nodes in relation $r$ space. The scoring function $g$ is thus given by
\begin{equation}
    \begin{split}
        g(h,r,t)=||W_re_h + e_r-W_re_t||
    \end{split}
\end{equation}
We optimize a separate BPR criterion powered KG scoring objective:
\begin{equation}
    \begin{split}
        \mathcal{L}_{KG}=\sum\limits{-\log\sigma \Big(g(h,r,t^\prime) - g(h,r,t)\Big)}
    \end{split}
\end{equation}
Combine \Cref{eq:multi-mle}:
\begin{equation*}
    \mathcal{L}^\prime = \mathcal{L} + \mathcal{L}_{KG}
\end{equation*}

%% file: text/experiment.tex
\section{Experiment}\label{sec:exp}
In this section, we experimentally investigate the performance of the proposed framework on two real world multi-behavior datasets. Specifically, we aim to answer the following research questions:
%\begin{itemize}\setlength{\itemindent}{-2em}
\begin{enumerate}[label ={}, leftmargin=0cm]
    \item \textbf{(RQ1)}: How do our proposed models: HMGN-intra and HMGN-inter perform compared to the SOTA baselines \citep{xia2021knowledge,chen2021graph} in terms of target-behavior prediction?
    \item \textbf{(RQ2)}: How does the proposed hierarchical multi-task learning perform compared to single-task learning in terms of target-behavior prediction? And how does it perform compared to other multi-task learning in terms of multi-behavior predictions?
    \item \textbf{(RQ3)}: How do our models perform when trained on sampled sub-graph vs non-sampling full graph?
    \item \textbf{(RQ4)}: How would different gadgets (temporal encoding, knowledge-graph) contribute to the performance?
\end{enumerate}
%\end{itemize}

\subsection{Datasets}
We use two publicly available multi-behavior e-commerce datasets: $Taobao$\footnote{\url{https://tianchi.aliyun.com/dataset/649}} (with \textit{view}, \textit{add-to-cart}, \textit{buy} , \textit{favor} four type behaviors) and $RetailRocket$\footnote{\url{https://www.kaggle.com/datasets/retailrocket/ecommerce-dataset}} (with \textit{view}, \textit{add-to-cart}, \textit{buy} three type behaviors). Works like \citep{xia2021knowledge} experiment with popular datasets like $Movielens$ and $Yelp$ by partitioning the user-item rating into different tiers to define target (like) and auxiliary behavior (dislike, neutral). However, in practical circumstances, a user cannot exhibit these behaviors simultaneously. Therefore, we refrain from using them for our experiments.

\subsubsection{Dataset Processing}
Although the \textit{leave-one-out} split strategy is widely adopted in academic research works \citep{lightgcn,xia2021knowledge,kgat,chen2021graph}, this is impractical in a real-world e-commerce scenario \citep{split}. 
To promote a rigorous and practical evaluation  \citep{diff}, we take the temporal split approach \citep{split}. Specifically, for $Taobao$ dataset, data from Nov-25-2017 to Dec-01-2017 is treated as training set, data in Dec-02-2017 is the validation set and Dec-03-2017 is the test set. Similarly in the $RetailRocket$ dataset, we split data according to the following timelines: May-03-2015 to Aug-15-2015 as a training set, Aug-16-2015 to Sep-01-2015 as the validation set and Sep-01-2015 to Sep-15-2015 as a test set.
For reproducibility, we would share the details of the data processing as well as the models in the code.
\subsubsection{Dataset Statistics}
We present the processed dataset statistic in \Cref{tab:dataset}
\begin{table}[]
\caption{Dataset statistic}
\label{tab:dataset}
\resizebox{\columnwidth}{!}{%
\begin{tabular}{|c|c|c|c|c|c|}
\hline
datasets     & \# user & \# item & \# interactions & \# categories & period \\ \hline
Taobao       & 48381       & 39369            &2018818                & 2511             & 11-25-2017 $\sim$ 12-03-2017     \\ \hline
RetailRocket & 14713       & 29286           & 207125               &  232           & 05-03-2015 $\sim$ 09-15-2015      \\ \hline
\end{tabular}
}
\end{table}
\subsection{Experimental Settings}
\subsubsection{Evaluation Metrics}
Same as \citep{vaecf}, we take commonly used $Recall@K$ and $NDCG@K$ ($K=10,50, 100$) as our metrics, specifically for a individual user $u$:
\begin{equation}\label{eq:metric}
    \begin{split}
        Recall@K (u) &=\frac{\sum\limits_{i\in I^{test}_u}\delta(R(i)\le K)}{min(K, |I^{test}_u|)}\\
        DCG@K (u) &=\sum\limits_{i\in I^{test}_u}\frac{\delta(R(i)\le K)}{\log( R(i) + 1)}\\
    \end{split}
\end{equation}
and $NDCG@K$ is the normalization of $DCG@K$. Here $\delta$ is an indicator function and $R(i)$ is the rank of item $i$ \citep{krichene@kdd20,li@kdd20,dong@aaai21,dong@aaai23}. $I^{test}_u$ is the set of positive items for user $u$ in test set. It is worth noting that in some special cases when $|I^{test}_u|> max(K_1, K_2)$ and  $K_1\ge K_2$., $Recall@K_1\le Recall@K_2$ could happen. This is because that numerator wouldn't change much (from $K_1$ to $K_2$), while the denominator could dominate the value.

\begin{table*}[t!]
\caption{Performance of the HMGN models w.r.t baselines on Taobao dataset. Mean results are present by repeating 5 times.}
\label{tab:taobao-main}
\resizebox{\textwidth}{!}{%
\begin{tabular}{|c|c|c|c|c|c|c|c|}
\hline
                                 & Model      & NDCG@10         & NDCG@50         & NDCG@100        & RECALL@10       & RECALL@50       & RECALL@100      \\ \hline
\multirow{3}{*}{Single-Behavior} & itemKNN    & 0.0052          & 0.0074          & 0.0087          & 0.0087          & 0.0178          & 0.0249          \\ \cline{2-8} 
                                 & BPR        & 0.0028          & 0.0049          & 0.0060          & 0.0049          & 0.0134          & 0.0195          \\ \cline{2-8} 
                                 & LightGCN   & 0.0065          & 0.0103          & 0.0123          & 0.0116          & 0.0269          & 0.0376          \\ \hline
\multirow{4}{*}{Multi-Behavior}  & LightGCN-M & 0.0079          & 0.0150          & 0.0188          & 0.0139          & 0.0424          & 0.0626          \\ \cline{2-8} 
                                 & KHGT       & 0.0022          & 0.0046          & 0.0063          & 0.0040          & 0.0144          & 0.0243          \\ \cline{2-8} 
                                 & KGAT       & 0.0093          & 0.0162          & 0.0209          & 0.0163          & 0.0444          & 0.0700          \\ \cline{2-8} 
                                 & GHCF       & 0.0170          & 0.0256          & 0.0304          & 0.0271          & 0.0621          & 0.0885          \\ \hline
\multirow{2}{*}{Proposed Model}  & HMGN-inter & 0.0156          & 0.0254          & 0.0303          & 0.025           & 0.0649          & 0.0917   \\ \cline{2-8} 
                                 & HMGN-intra & 
\textbf{0.0354} & \textbf{0.0453} & \textbf{0.0499} & \textbf{0.0474} & \textbf{0.0874} & \textbf{0.1124} \\ \hline
\end{tabular}
}
\end{table*}

\begin{table*}[]
\caption{Performance of the HMGN models w.r.t baselines on RetailRocket dataset. Mean results are present by repeating 5 times.}
\label{tab:retail-main}
\resizebox{\textwidth}{!}{%
\begin{tabular}{|c|c|c|c|c|c|c|c|}
\hline
                                 & Model      & NDCG@10         & NDCG@50         & NDCG@100        & RECALL@10       & RECALL@50       & RECALL@100      \\ \hline
\multirow{3}{*}{Single-Behavior} & itemKNN    & 0.0012          & 0.0017          & 0.0027          & 0.0005          & 0.0057          & 0.0111          \\ \cline{2-8} 
                                 & BPR        & 0.0078          & 0.0072          & 0.0094          & 0.0163          & 0.0160          & 0.0289          \\ \cline{2-8} 
                                 & LightGCN   & 0.0103          & 0.0090          & 0.0088          & 0.0144          & 0.0132          & 0.0131          \\ \hline
\multirow{4}{*}{Multi-Behavior}  & LightGCN-M & 0.0559          & 0.0660          & 0.0681          & 0.0844          & 0.1217          & 0.1295          \\ \cline{2-8} 
                                 & KHGT       & 0.0037          & 0.0079          & 0.0104          & 0.0111          & 0.0284          & 0.0426          \\ \cline{2-8} 
                                 & KGAT       & 0.0645          & 0.0749          & 0.0794          & 0.0743          & 0.1162          & 0.1407          \\ \cline{2-8} 
                                 & GHCF       & 0.0100          & 0.0135          & 0.0162          & 0.0144          & 0.0341          & 0.0468          \\ \hline
\multirow{2}{*}{Proposed Model}  &  HMGN-inter & 0.0710          & 0.0805          & 0.0857          & 0.0909          & 0.1267          & \textbf{0.1553} \\ \cline{2-8} 
                                 & HMGN-intra & \textbf{0.0957} & \textbf{0.1044} & \textbf{0.1070} & \textbf{0.1070} & \textbf{0.1398} & {0.1517} \\ \hline
\end{tabular}
}
\end{table*}

\subsubsection{Baselines}
We compare the proposed model with several influential recommendation models including both single-behavior and multi-behavior ones.

\noindent\textbf{Single-behavior Models:}
\begin{itemize}\setlength{\itemindent}{-1em}
\item \textbf{itemKNN} \citep{itemknn} A classical and robust neighborhood method.
    \item \textbf{BPR} \citep{rendle2009bpr} is one of the most popular methods in recommendation which learns to optimize pair-wise objectives.
    \item \textbf{LightGCN} \citep{lightgcn} is a graph-based model which simplifies the framework of GCN for recommendation by removing feature transformation and nonlinear activation. 
\end{itemize}
\noindent\textbf{Multi-behavior Models:}
\begin{itemize}\setlength{\itemindent}{-1em}
\item \textbf{LightGCN-M} We extend LightGCN for multi-task optimization by integrating multi-behaviors during the modeling stage.
    \item \textbf{KHGT} \citep{xia2021knowledge} is a SOTA model that targets on multi-behavior recommendation task enhanced with knowledge and temporal information.
    \item \textbf{KGAT} \citep{kgat} attention-based graph neural network that incorporates the item meta-data. Here, we extend and optimize it with multi-task learning.
    \item \textbf{GHCF} \citep{chen2021graph} a SOTA model particularly for multi-behavior collaborative filtering without sampling any negative items by optimizing a mean square error (MSE) loss.
\end{itemize}
\subsection{Performance Comparison (RQ1)}

\Cref{tab:taobao-main,tab:retail-main} show the performance of all models on $Taobao$ dataset and $RetailRocket$ dataset respectively. We highlighted the best performance in each column. Here are some main observations:
\begin{itemize}\setlength{\itemindent}{-1em}
    \item Our proposed model HMGN-intra can outperform all baselines significantly and consistently. HMGN-intra improves performance by 35\% (RetailRocket) and by 64\% (Taobao) as compared to the best baseline in terms of $NDCG@100$. By leveraging the attention mechanism, HMGN-intra is able to capture the multi-behavior information that helps target behavior prediction.
    \item HMGN-intra is generally better than HMGN-inter, indicating that behavior-specific learning is more effective than cross-behavior learning for multi-behavior recommendations. In the HMGN-intra model, all information propagates first in the isolated single-behavior graph and then gathers together. Considering the multi-task objectives, these would better help explain and contribute to each individual behavior prediction, leading to a superior performance.
    
    \item most all the multi-behavior models exhibit better performance compared to single-behavior ones, this emphasizes the importance of multi-behavior utilization.
\end{itemize}

\subsection{HBPR Multi-task Optimization (Q2)}
\begin{figure}
    \centering
    \includegraphics[width=0.9\linewidth]{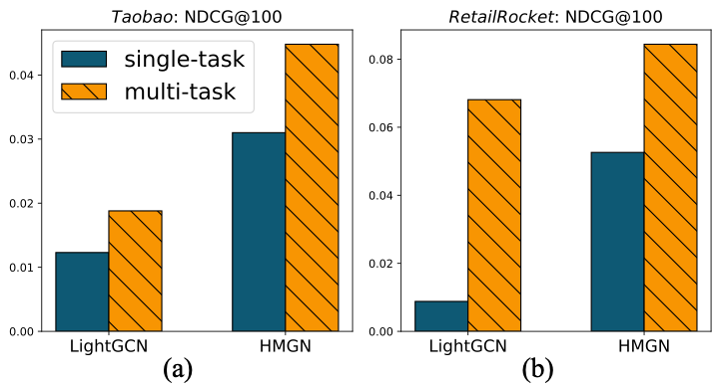}
    \vspace{-5pt}
    \caption{Single-task learning $vs$ multi-task learning.}
    \label{fig:single-multi}
\end{figure}

We empirically testify the HBPR based multi-task optimization framework from two perspectives:

\noindent\textbf{1. Speciality.}
Comparing the performance of LightGCN with its counterpart LightGCN-M, and base model (HMGN-intra optimized by HBPR) with its counterpart $wo. multi-task$ in \Cref{fig:single-multi}, we would see that the multi-task optimization is crucial for multi-behavior recommender system. In other words, even if the end goal of the model is to predict one target behavior, learning to explicitly optimize auxiliary behaviors in model objective significantly improves the model metrics.

\begin{figure}
    \centering
    \includegraphics[width=0.9\linewidth]{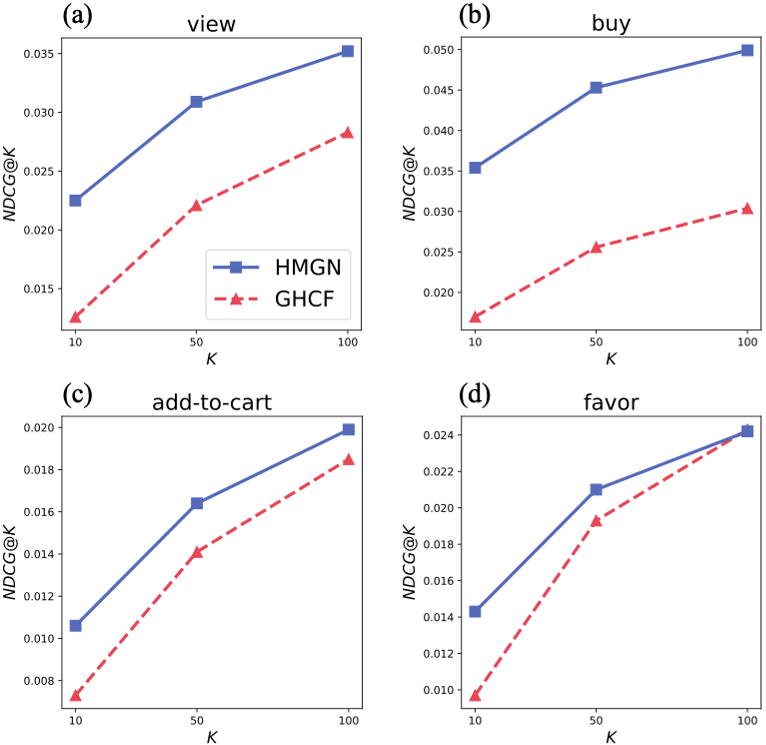}
    \caption{Performance comparison (HMGN-intra $vs$ GHCF) on multi-behavior predictions for $Taobao$ dataset, measured by $NDCG@K (K = 10, 50, 100)$.}
    \label{fig:multi-behaviors}
\end{figure}

\noindent\textbf{2. Expressiveness.}
We also compare our proposed model with GHCF on the ability to predict other auxiliary behaviors. \Cref{fig:multi-behaviors} indicate our method consistently outperforms GHCF on all behaviors prediction in terms of $NDCG$ metric.

\subsection{Sub-graph Sampling (Q3)}
To test the scalability of our graph, we utilized the algorithms that are proposed in \Cref{subsec:sub-sampling}. We assign different sampling size and obtain the performance for each setting in \Cref{tab:sub-graph-size}. (Since the sub-graph size is dependent on the sampling size, we repeat this experiment 100 times to compute an average sub-graph size.) Note that for both the datasets, a sub-graph size of around $20K$, can already get pretty good results compared to the full-size graph. It is also worth noting that sub-graph size need not grow as the size of the dataset increases. Overall this effective sub-graph sampling algorithm enables us to execute the model on a large datasets.

\begin{table}[]
\caption{Performance of HMGN-intra w.r.t sampled sub-graph size.}
\label{tab:sub-graph-size}
\resizebox{\columnwidth}{!}{%
\begin{tabular}{|c|c|c|c|c|c|c|}
\hline
\multirow{3}{*}{Taobao}                                                        & sub-graph size   & 20K    & 25K    & 30K    & 40K    & 88K (entire) \\ \cline{2-7} 
                                                                               & NDCG@100         & 0.0387 & 0.0462 & 0.0475 & 0.0483 & 0.0499               \\ \cline{2-7} 
                                                                               & Recall@100       & 0.0918 & 0.1080  & 0.1083 & 0.1109 & 0.1124               \\ \hline
\multirow{3}{*}{\begin{tabular}[c]{@{}c@{}}Retail-\\      Rocket\end{tabular}} & sub-graph   size & 11K    & 14K    & 19K    & 24K    & 44K   (entire) \\ \cline{2-7} 
                                                                               & NDCG@100         & 0.0904 & 0.0943 & 0.0948 & 0.0961 & 0.1070                \\ \cline{2-7} 
                                                                               & Recall@100       & 0.1494 & 0.1450  & 0.1539 & 0.1508 & 0.1517               \\ \hline
\end{tabular}
}
\end{table}

\subsection{Study of Graph Enhancement (Q4)}
\begin{figure}
    \centering
    \includegraphics[width=0.9\linewidth]{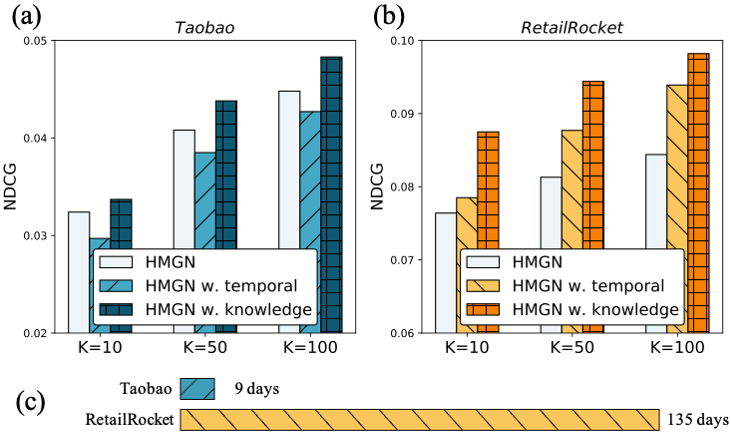}
    \caption{Temporal and knowledge effect. Temporal encoding benefits the HMGN model on the dataset with long term data ($RetailRocket$ with 135 days data points).}
    \label{fig:temporal_knowledge}
\end{figure}

In this section, we investigate how different graph enhancing gadgets affect our best performing HMGN-intra model. From \Cref{fig:temporal_knowledge}, we have two key takeaways:
\begin{itemize}\setlength{\itemindent}{-1em}
    \item In $Taobao$ dataset (\Cref{fig:temporal_knowledge}(a)), HMGN-intra with temporal encoding performs worse than that without temporal information whereas in $RetailRocket$ dataset (\Cref{fig:temporal_knowledge}(b)) HMGN-intra with temporal encoding included gets better performance than that without temporal encoding. We think this contradictory effect is caused by the duration of the data (\Cref{fig:temporal_knowledge}(c)). In $Taobao$, all data is collected within 9 days which all can be considered as "recent" events. In this case, the fusion of temporal encoding would behave just like a noise injection that decreases the performance. In $RetailRocket$, a long period dataset, the data lasts 3.5 months, which would enable the temporal encoding to take effect.
    \item Also indicated in \Cref{fig:temporal_knowledge}, the knowledge information specifically item metadata here, is particularly beneficial to both two e-commerce datasets.
\end{itemize}

%% file: text/related.tex
\section{Related Works}\label{sec:related}
Our work builds upon a few recent advancements. We list related works in this section and compare the resemblance and discrepancy between them and ours.

\subsection{Multi-behavior Recommender System} 
The multi-behavior recommendation is an emerging field that utilizes user-item multi-behavior interactions to predict target behavior. \citep{DMBGN,chen2021graph,xia2021knowledge,mbgcn,mbgmn,nmtr,multiview-cl,mbs}. MBGCN \citep{mbgcn} utilizes the multi-behavior information through a graph convolutional neural network. KHGT \citep{xia2021knowledge} is a SOTA model that incorporates knowledge and temporal information for the multi-behavior recommendation task enhancing model performance. GHCF \citep{chen2021graph} achieves SOTA results through multi-behavior collaborative filtering. 
There are also some sequential-based multi-behavior recommendations. MMCLR \citep{multiview-cl} learns sequence view and graph view multi-behavior model. MBHT \cite{mbht} proposes a sequential recommendation framework that enables the dependencies of both short-term and long-term multi-behavior information.

\subsection{Graph Enrichment and Augmentation}
With the ability to learn high-order topological information, Graph Neural Network (GNN) \cite{xiang@scalable,Xu2020Inductive} has achieved significant success in Recommender System \citep{lightgcn,NGCF19,peng@2022sigir}. Various techniques like optimizing Contrastive Learning (CL) \citep{sslrec}, leveraging metadata or Knowledge Graph (KG) \citep{kgat}, incorporating temporal information \citep{Xu2020Inductive}, sampling sub-graph \citep{pinsage} for scalability, are exhibiting advantages to exploit robust and powerful GNN. We reason such techniques as graph enrichment and augmentation. KGAT \citep{kgat} integrates the KG and attention mechanism for single behavior recommendations systems where the objectives are consist of with both KG \citep{transr} and collaborative filtering optimization. SGL \citep{sgl} generates multiple views of graphs by node and edge drop and applies InfoNCE \citep{infonce} onto that. SimGCL \citep{simgcl} 

\subsection{Resemblance and Discrepancy}

In terms of architecture, GNMR \citep{mb-cross}, KHGT \citep{xia2021knowledge} share the multi-behavior graph attention backbone with HMGN-intra, while HMGN-intra is lighter with fewer learning parameters and a simple representation combination. When handling knowledge information, instead of allowing information propagation in a hybrid graph like KHGT and KGAT \citep{kgat}, we only optimize for an independent KG objective to enable efficiency. When it comes to objectives, while the majority of multi-behavior models only learn to optimize and predict target behaviors \citep{xia2021knowledge,mbgcn}, there are few works that are similar to this. NMTR \citep{nmtr} treats all observed behavior as (weighted) positive and sample negative from non-interacted items as negative, without any hierarchical relation between behaviors. Similar to us, \citep{multiview-cl} defines the behavior priority rank but applies the BPR to two behaviors in adjacent order. GHCF \citep{chen2021graph} is the only existing model that establishes predictions for all behaviors. They optimize a mean square error (MSE) loss without sampling any negative items while we utilize HBPR considering the hierarchical relations between behaviors. Besides, GHCF is a GCN-based model, compared to our proposed HMGN model which is an attention-based GNN, allowing temporal encoding.

%% file: text/conclusion.tex
\section{Conclusion}\label{sec:conclusion}
In this paper, we devised two new graph attention-based frameworks called HMGN-intra and HMGN-inter for multi-behavior recommender systems. We discover that it is crucial for multi-behavior systems to learn via multi-task objectives, aka, optimizing for all behaviors instead of target behaviors. To this end, we propose a hierarchical Bayesian Personalized Ranking optimization criterion. We enable our model with the ability to capture preference generalization as well as behavior specialization and to predict all types of behaviors. Further, we provide a unified and comprehensive strategy for multi-behavior methods. Specifically, we enhance our model with multiple techniques like temporal encoding and metadata information. To scale up the model, we also extend the sub-graph sampling to the multi-behavior scenarios. Extensive empirical analyses indicate our proposed model outperforms the baselines significantly. 